\documentstyle[amssymb,prl,aps,twocolumn,epsfig]{revtex}

\begin{document}
\draft
%
%
%
%
%
%
%
\twocolumn[\hsize\textwidth\columnwidth\hsize\csname @twocolumnfalse\endcsname

\title{Correlations due to localization in quantum eigenfunctions of disordered microwave cavities}
\author{Prabhakar  Pradhan and S. Sridhar}
\address{Department of Physics \\
Northeastern University\\
 Boston, Massachusetts 02115}

\date{\today}

\maketitle

\begin{abstract}
Non-universal correlations due to localization are observed in statistical properties of experimental eigenfunctions of quantum
chaotic and disordered microwave cavities.
Varying energy$\ E$ and mean free path $l$ enable us to
experimentally tune from localized to delocalized states. Large level-to-level Inverse Participation
 Ratio (IPR $I_{2}$) fluctuations are observed for the disordered billiards, whose distribution is strongly asymmetric about
 $\langle I_{2}\rangle $. The density auto-correlations of eigenfunctions are shown to decay
exponentially and the decay lengths  are  experimentally  determined.
 All the results are
 quantitatively consistent with calculations based
 upon  nonlinear $\sigma -$models.
\end{abstract}
\vspace{.3cm}

\vskip.5cm]
%

The universal properties of quantum chaotic systems have been extensively
studied in terms of their eigenvalue and eigenfunction statistics \cite
{gutzwiller}. In Random Matrix Theory (RMT), the Gaussian distribution of
eigenfunction amplitudes $\psi (\vec{r})$ leads to the universal
Porter-Thomas (PT) distribution for the densities $|\psi |^{2}$, which has
been observed in microwave cavities \cite{kudrolli95a} as well as other
systems. However, non-universality has important manifestations, for
instance due to periodic orbits which lead to scars in eigenfunctions, and
localization arising from quantum interference in diffusion. While there
have been many theoretical treatments, from semiclassical periodic orbit
theories \cite{gutzwiller} to nonlinear sigma models \cite{Efetovbook},
there have been few experimental studies of eigenfunctions because of their
lack of accessibility.

In this paper, we present several striking manifestations of localization in
experimental eigenfunctions of disordered microwave billiard cavities.
Localization due to boundary or impurity scattering results in correlations
that affect statistics and spatial correlations of the eigenfunctions in
several measures, leading to deviations from their universal values for
quantum chaotic systems. The moments of the density distribution, $%
I_{n}=\int |\Psi (\bar{r})|^{2n}d^{3}r$, in particular the Inverse
Participation Ratio $I_{2}$ $(IPR)$, and their distributions $%
P_{I_{n}}(I_{n})$, are important measures of the properties of the chaotic
and disordered eigenfunctions. In chaotic billiards, $I_{2}$ has a mean
value $\langle I_{2}\rangle $ close to that of the universal $2$-dimensional 
$(2D)$ limiting value of $3.0$, with small level-to-level fluctuations $%
\delta I_{2}\ll \langle I_{2}\rangle $, resulting in a nearly symmetric
distribution about $\langle I_{2}\rangle $. In disordered billiards not only
is the mean value $\langle I_{2}\rangle \gg 3.0$, but the fluctuations are
also much greater $\delta I_{2}\sim \langle I_{2}\rangle $. The IPR
distribution for the disordered billiards is strongly asymmetric about $%
\langle I_{2}\rangle $, and is quantitatively consistent with the
calculations based upon the nonlinear $\sigma -$models of supersymmetry,
parametrized by a conductivity $g$ \cite{prigodin98,mirlin99}. Spatial
correlations are studied in terms of the density auto-correlation $%
\left\langle |\Psi (r)|^{2}|\Psi (r^{^{\prime }})|^{2}\right\rangle $ and
are shown to die out more rapidly in the disordered billiards compared with
the chaotic geometries with a characteristic decay length given by the mean
free path $l$. Here again the data are in quantitative agreement with the $%
\sigma -$model calculations. Our results represent the first quantitative
comparison of experiments and theory.

The experiments were carried out using thin $($ height $d$ $<6$ mm$)$
cavities, whose cross-sections can be shaped in essentially arbitrary
geometries. For these two dimensional cavities, the operational wave
equation is $(\nabla ^{2}\,+\,k^{2})\Psi =0$, where $\psi =E_{z}$ is the
microwave electric field. Similar microwave experiments, which exploit this
QM-EM mapping, have been used to study quantum chaos in closed \cite
{sridhar91,microwavechaos} and open systems \cite{lu99}. Eigenfunctions $%
|\psi (r)|^{2}$ were directly measured using a cavity perturbation technique 
\cite{sridhar91}. Localization effects were observed by fabricating
billiards in which $1cm$ circular tiles were placed in a $44cm$ x $21.8cm$
rectangle at random locations (Fig.1) to act as hard scatterers. (The
locations were generated using a random number generator and the tiles
placed manually). Several realizations of the disordered geometry were
experimentally studied, by varying the density of scatterers from $12$ to $71
$, thus varying mean free path $l\sim 8.9cm$ to $3.6cm$. Earlier experiments 
\cite{kudrolli95a} had shown that these disordered geometries are an
excellent experimental realization of the classic problem of an electron in
a 2-D disordered potential. Subsequent to this work, there have been
important theoretical developments \cite{falko94,prigodin98,mirlin99}, some
motivated by our earlier experiments.

The experimental eigenfunctions directly demonstrate the trend toward
decreasing localization from disordered-chaotic-integrable as seen in Fig.%
\ref{fig1}, which shows representative experimental eigenfunctions, along
with their corresponding IPR $I_{2}$, for several billiards. The strongly
localized state Fig.\ref{fig1}(a) at $f=3.04GHz$ of the disordered billiard
with $N=36$ scatterers has very large $I_{2}=13.42$. In contrast the more
delocalized state at higher frequency $f=7.33GHz$ (Fig.\ref{fig1}(b))
explores almost all the available coordinate space similar to chaotic
cavities and has a
smaller $I_{2}=4.06$. For the chaotic Sinai stadium,
typical values of the IPR are around $3.0$ ($I_{2}=3.01$ for this
eigenfunction Fig.\ref{fig1}(c)) while for the integrable rectangle billiard
Fig.\ref{fig1}(d) $IPR\,$ $I_{2}=2.25$ for all eigenfunctions. Fig.\ref{fig1}
demonstrates a key advantage of our experiments, which is that by varying $l$
and wavevector $k$ we are able to access a wide range of the disorder
strength $kl$, from strongly localized states for $kl<1$ to delocalized
states with $kl\gg 1 $.

Eigenfunctions such as in Fig.\ref{fig1} were then analyzed in terms of $%
I_{2}$ and $P_{I_{2}}(I_{2})$. In the following, for convenience, we use the
notation {\bf \- }$I_{2}{\bf =}\int |\Psi (r)|^{4}dv=\int u^2dv,w=(I_{2}-3)/6$
, $u=|\Psi (r)|^{2}$, and $I_{1}{\bf =}\int |\Psi (r)|^{2}dv=1$, and the
integral is over the volume $v$ (area in $2D$ ). Nearly 250 eigenfunctions
were analyzed each containing more than 3200 eigenfunction values.

Fig.\ref{fig2} shows level-to-level variations $I_{2}(E)$ for the Sinai
stadium. Here $I_{2}$ is mainly clustered around a mean value of $%
\left\langle I_{2}\right\rangle =3.0$, with relatively small level-to-level
fluctuations. The IPR distribution $P_{I_{2}}(I_{2})$ of the chaotic
Sinai-stadium billiard is shown in Fig.\ref{fig3}(top). $P_{I_{2}}(I_{2})$
is seen to be a nearly symmetric distribution with a small width $%
I_{2}-\left\langle I_{2}\right\rangle \ll \left\langle I_{2}\right\rangle $.

Ref.\cite{kudrolli95a} demonstrated that the Sinai-stadium billiard data
obey the universal Porter-Thomas (P-T) distribution $P_{u}(u)=(2\pi
u)^{-1/2}\exp (-u/2)$, with $u=|\Psi |^{2}$ to remarkable degree, while
deviations from P-T were demonstrated due to localization in the disordered
billiards. The IPR for P-T distribution can be immediately obtained $%
I_{2}=\int_{0}^{\infty }u^{2}P_{u}(u)du=3.0$, which is a universal value.
Note that there are no fluctuations expected about this universal value in 
RMT, i.e., $P_{I_{2}}(I_{2})$ is a $\delta $-function at $I_{2}=3$ \cite
{srednicki}. Clearly boundary scattering on the system length scale $R$
leads to non-universal correlations (e.g. from periodic orbits leading to
scars in the wavefunctions). This breaks the assumption in RMT of Gaussian
fluctuations of the eigenfunction amplitude, and in turn leads to
fluctuations in the distribution $P_{I_{2}}(I_{2})$ (although of narrow
width) observed in Fig.\ref{fig3}(top).

\smallskip Even more strikingly, the IPRs of the disordered billiards shown
in Fig.\ref{fig2} display a remarkably large spectrum of level-to-level
fluctuations (Fig.\ref{fig2}). For small $f$, when $\lambda >l$, strong
localization leads to large values of $I_{2}$ in the disordered cavity,
which can be as high as $\sim 20$. It is worth noting that the density
distributions $P_{u}(u)\,$of the eigenfunctions deviate strongly from the
P-T distribution and are consistent with the large IPR values.(In this paper
we have focussed on the billiard with $l=5.1cm.$, but similar results were
obtained for other billiards and will be presented in a larger publication.)

As $f$ is increased (or $\lambda $ is decreased), the IPR progressively
decreases, approaching the universal limiting value of $3$, as shown in Fig.%
\ref{fig2}. The corresponding distribution $P_{I_{2}}(I_{2})$ shown in Fig.%
\ref{fig3}(bottom) is strikingly different from the Sinai-stadium case, in
that, it is strongly asymmetric, reflecting the very large $I_{2}\gg
\left\langle I_{2}\right\rangle $ values that are present, and is strongly
non-Gaussian. The IPR fluctuations in Fig.\ref{fig2} are closely
similar to the famous universal conductance fluctuations of a mesoscopic
system.

Recently several theoretical studies have calculated the IPR distribution
based on the supersymmetry method \cite{prigodin98,mirlin99}. For a
mesoscopic system the IPR distribution depends on the conductivity $g$ of
the system, defined as $g=\ln (R/l)/\langle w\rangle $ , where $R$ is the
system size and $l$ is the mean free path and $\langle ..\rangle $ is the
realization average for a fixed ``disorder strength'' $2kl$. The resulting
probability 
distribution $P(I_{2})$ for $I_{2}<\langle I_{2}\rangle =3$ is 
\cite{prigodin98}:
\begin{equation}
P_{I_{2}}(I_{2})=C_{1}\frac{g}{2}\exp [-\frac{g}{6}(I_{2}-\langle
I_{2}\rangle )-\frac{\pi }{2}e^{-\frac{g}{3}(I_{2}-\langle I_{2}\rangle )}],
\end{equation}

and the corresponding distribution for $I_{2}\gg \langle I_{2}\rangle $ : 
\begin{equation}
P_{I_{2}}(I_{2})=C_{2}\sqrt{\frac{g}{I_{2}}}\exp (-\frac{\pi }{6}gI_{2})
\end{equation}

where $C_{1\text{ }}$ and $C_{2}$ are normalization constants.

The solid line in Fig.\ref{fig3}(bottom) represents Eq.(2)  for $I_{2}\gg
\langle I_{2}\rangle $ using a small conductivity $g=1.0$ and is seen to
describe the data very well. Another way of presenting the comparison with
Eq.(2) is by using a scaled variable $q=gI_{2}$ whence the distribution changes
to a Porter-Thomas distribution in $q$. The system is within the weak
disorder limit, i.e. the random phase approximation still valid, and $g$
depends weakly on the disorder factor $2kl$ \cite{pradhan}. Now rescaling
the $I_{2}$ with $g$ we obtain the distribution $P_{q}(q)$ (unnormalized)
and plot $\ln [P_{q}(q)\sqrt{q}]$ vs $q$ in Fig.\ref{fig4}. This shows a
straight line which implies a good agreement of theory and experiment for
IPR distribution in a moderate disordered region where $I_{2}\gg \langle
I_{2}\rangle $. This is the first experimental observation of the asymmetric
distribution predicted by Prigodin and Altshuler \cite{prigodin98}.

We now return to the case of the Sinai-stadium. Although a formulation in
terms of a ballistic sigma model has been presented for chaotic cavities\cite
{ballisticsigma}, it is not simply amenable to experimental comparison.
Instead we use Eq.1 with the assumption that since $l\approx R,$ a suitably
large conductivity $(g>>1)$ can be used. In our experimental case, $g\approx
7.8$ matches Eq.1 not only for $I_{2}<\langle I_{2}\rangle $, but also for $%
I_{2}>\langle I_{2}\rangle ,$ as shown in Fig.3(top). The nearly symmetric
distribution can be understood since the fluctuations arise from
correlations at the scale of the system length $R$, which is the only length
scale in the problem. The values of $g$ corresponding to the disordered
billiards and the Sinai-stadium billiard are consistent with the $%
\left\langle w\right\rangle $ values and the mean free paths.

We now show that the experimental data also directly demonstrate the decay
of spatial correlations $\left\langle \Psi ^{2}(r)\Psi ^{2}(r^{\prime
})\right\rangle $ of the eigenfunctions and obtain the decay length, which
corresponds to the classical mean free path. To calculate the correlation
with arbitrary disorder strength $2kl$, defining $K(r)=|%
\mathop{\rm Im}%
G(r^{\prime })|^{2}/(\pi \nu )^{2}$ , where $G(|r-r^{^{\prime
}}|)=<r|(E-H)^{-1}|r^{\prime }>$ is the Green function of the disordered
system Hamiltonian, then it can be shown \cite{prigodin94} that $K(r)=|\frac{%
1}{\pi }\int_{-\infty }^{\infty }\frac{1}{1+y^{2}}J_{0}[kr(1+\frac{1}{2kl}%
y)]dy|^{2}$.

For the chaotic or ballistic limit $(2kl>>1),$ the result for Gaussian
fluctuations is $\left\langle \Psi ^{2}(r)\Psi ^{2}(r^{\prime
})\right\rangle \simeq 1+(I_{2}-1)J_{0}^{2}(r-r^{\prime }),$ where $J_{0}$
is the first order Bessel function . The correlation for a moderate
disordered case with correct limits can be derived repeating the
calculations of Ref:\cite{prigodin95}

\begin{eqnarray}
\left\langle \Psi ^{2}(r)\Psi ^{2}(r^{\prime })\right\rangle &\simeq
&1+(I_{2}-1)K(k|r-r^{\prime }|) \\
&\simeq &1+(I_{2}-1)J_{0}^{2}(k|r-r^{\prime }|)e^{-\frac{k|r-r^{^{\prime }}|%
}{kl}}  
\end{eqnarray}

These are valid when $r-r^{\prime }\precsim l$. We have solved $%
K(k|r-r^{^{\prime }}|)$ numerically. In Fig.\ref{fig5}, we plot the average
correlation derived from experimental data, numerical calculations of
Eq.(3), and analytical expression of Eq.(4) for different disorder strengths 
$2kl.$ For the Sinai billiards, the average correlation starts at $3$ and
tend to $1$ via Friedel oscillations, consistent with Eq.(4) with $2kl=37$,
which is very large and hence the result is close to that of the universal
dependence given by $kl=\infty $. For disordered billiards, the
auto-correlation is very large ($\sim I_{2}$) for short lengths, i.e. around 
$|r-r^{\prime }|\approx 0$ due to localization, but the auto-correlation
decays with a decay length scale, responsible for fast fall, and it should
go to zero at large $k|r-r^{\prime }|$. The experimental data shows good
agreement with the numerical and analytical calculations for moderate
disorder, as shown in Fig.\ref{fig5} for values $2kl=12$ and $7$. These
values are in excellent agreement with the corresponding mean free paths $%
(l=3.6$, $5.1$, $5.9$ and $8.9)$ obtained by directly considering the number
of scatterers $N$, so that $l\sim \sqrt{ab/N}$ and $a$ and $b$ are the sides
of the enclosing rectangle. Thus our analysis directly demonstrates
localization and yields a quantitative measure in terms of the correlation
decay length.

Returning to the level-to-level $I_{2}$ data in Fig.\ref{fig2} we note that
they can be viewed as a form of localization-delocalization transition,
tuned by the frequency $f$ ! Representing the IPR as $I_{2}(E)=I_{2,sm}(E)+%
\delta I_{2}(E)$, i.e. a smooth part $I_{2,sm}(E)$ plus fluctuations $\delta
I_{2}(E)$, we discuss the trends in $I_{2}(E)$. Calculations in Ref.\cite
{fyodorov97} based on infinite dimensional tight binding model shows that $%
I_{2}(E)$ diverges exponentially near the critical point $E_{c}$: $%
I_{2}(E)=I_{2}(E=\infty )\exp (A/|E-E_{c}|^{\mu })$, with $\mu =\frac{1}{2}$
, due to very strong correlations of the wave function near $E_{c}$. $%
I_{2}(E=\infty )$ will be obviously the asymptotic universal value $3,$ as
predicted by RMT in $2D$. Our experimental $IPR$ data is as large as $IPR$ $%
\sim $ $20$ (strongly localized), decaying to $\sim 4$ (weakly localized) at
high frequencies upto $10GHz.$ In the present case, the smallest scale of
the system $\sim 5cm$ sets a natural lower cut-off frequency $%
f_{c}=c/2l=3GHz $, so that there are no eigenstates for $E<$ $%
E_{c}=f_{c}^{2}.$ For $E\gg E_{c}$ expanding the above equation in first
order, we obtain $I_{2,sm}(E)\simeq 3+B/|E-E_{c}|^{\mu }$. In Fig.\ref{fig2}
we have plotted this expression with $\mu =0.5$ and $B=9.0$ (obtained by a
best fit) and compared with the experimental data. The above expression
captures the trend of the data. While an exact comparison with any
expression is difficult since the fluctuations $\delta I_{2}(E)>I_{2,sm}(E),$
the analysis illustrates that we are observing a frequency driven localized
to de-localized transition in a disordered medium in terms of the $IPR.$

We have shown that the IPR is an extremely valuable parameter to study
real-space localization in quantum eigenfunctions. We have demonstrated for
the first time a quantitative analysis of features well beyond universality
due to localization in experimental eigenfunctions. The observed IPR
distribution is strongly non-Gaussian due to the correlations induced by
scattering. The nonlinear sigma-model is in quantitative agreement with the
experimental data for moderate localization. Our work thus provide
experimental support for the approach based upon quantum diffusion in the
localization regime.

This work was supported by NSF-PHY-9722681. S.S. thanks the hospitality of
the Quantum Chaos Workshop at the\ Australian National University, Canberra,
where part of this work was carried out. We thank A. Kudrolli for useful
discussions.

\narrowtext%

\begin{figure}
\caption{(a,b) Experimental eigenfunctions of a disordered  billiard with 
$n = 36$ scatterers (noted by the black dots) (a) a strongly localized state, $f = 3.04GHz$,
$I_2 =13.42$  and (b) a delocalized state $f = 7.33 GHz$, $I_2=4.06$.
Eigenfunctions of   (c) Sinai-stadium billiard $I_2 = 3.01$,
(d) an integrable rectangle ($I_2 = 2.25$ for all states).
}
\label{fig1}
\end{figure}%

\begin{figure}
\caption{Large level-to-level fluctuations are observed in the IPR $I_2 (E)$ v.s. $E$ of the disordered billiard.
Also note the gradual trend towards 
a universal limiting value of 3.0 indicated by the solid line which represents a model described in the text.
The fluctuations of the Sinai-stadium are much smaller and are clustered around $< I_2 > = 3.0$.}
\label{fig2}
\end{figure}%

\begin{figure}
\caption{ IPR distribution $P_{I_2} (I_2 )$ of the disordered billiard (bottom) is strongly
 asymmetric and non-Gaussian.  
The distribution for the chaotic Sinai-stadium billiard (top)  is nearly symmetric about the mean
 value $3.0$. The lines represent  calculations based on the nonlinear
sigma-model.}
\label{fig3}
\end{figure}%

\begin{figure}
\caption{ The IPR distribution of Fig.3 replotted using the scaled variable $q=g I_2$ . 
The solid line is the Porter-Thomas distribution  $ln [ P_q(q)  q^{1/2}  ]$  vs  $q$
and agrees for large $I_2 >> <I_2>$.}
\label{fig4}
\end{figure}%

\begin{figure}
\caption{Density auto-correlation $\left\langle \Psi ^{2}(r)\Psi ^{2}(r^{\prime
})\right\rangle$ of  eigenfunctions of  the  Sinai-stadium and disordered billiards 
 with fixed disordered strength $2kl$. Experiment ( dotted lines), Eq.$(3)$ (dashed lines),
 and Eq.$(4)$   with $2kl$ =$7$, $12$ and $37$ (solid lines).}
\label{fig5}
\end{figure}%

\end{document}